\documentclass[12pt]{article} 

\usepackage{amsmath, amssymb, graphics}
\newcommand{\mathsym}[1]{{}}
\newcommand{\unicode}[1]{{}}

\title{The Lorentz Group with Dual-Translations and the Conformal Group}  
\author{{\it Richard Shurtleff~}\thanks{affiliation and mailing 
address: Department of Science, 
Wentworth Institute of Technology, 550 Huntington Avenue, 
Boston, MA,  02115, USA, telephone: (617) 989-4338, FAX:  617-989-4010, e-mail: shurtleffr@wit.edu}} 
\begin{document} 
%\AmS-\LaTeX
          
\maketitle 

\begin{abstract} 

For those finite-matrix representations of the Lorentz group of rotations/boosts with spin $(A,B)\oplus(C,D)$ that can also represent translations, two possible translation subgroups qualify. Of these two, one must be selected,  and one discarded, to represent the Poincar\'{e} group of rotations/boosts with translations in spacetime. Instead, let us discard the requirement that there be just one translation subgroup. With dual-translations, one gives up agreement with simple macroscopic observations of spacetime. Now the transformations of both possible translation subgroups combine with those of the Lorentz group. The resulting commutation relations require new transformations and generators to satisfy the linearity requirement of a Lie algebra. Special cases of spins are sought to restrict the influx of new transformations.  One finds that the Dirac 4-spinor formalism is the only viable solution. The slightly expanded group it represents is the conformal group with just one new transformation, scale change. It follows as a corollary that the Dirac 4-spinor formalism is the only matrix representation of the conformal group with spin $(A,B)\oplus(C,D).$

\vspace{0.5cm}

\end{abstract}
\pagebreak
\section{Introduction} \label{into}

The Lorentz group describes rotations and boosts in spacetime, transformations that preserve the scalar product of 4-vectors. Translations also preserve the scalar product of 4-vectors because, in conventional notation, the 4-vector coordinate interval $\delta x^{\mu}$ = $(x^{\mu}_{2} + a^{\mu}) - (x^{\mu}_{1} + a^{\mu})$ = $x^{\mu}_{2}  - x^{\mu}_{1} $ from event 1 to event 2 is unchanged by a displacement $a^{\mu}$ of the origin. The combined group of rotations, boosts and translations is the Poincar\'{e} group.

In quantum mechanics and elsewhere, one encounters finite dimensional, non-unitary matrix representations (reps) of the Lorentz group. In this paper the focus is on matrix reps, not differentiable reps defined on a multi-dimensional manifold like spacetime.

Some matrix reps of the Lorentz group can be extended to also represent translations.  Not all Lorentz reps qualify. Irreducible spin $(A,B)$ Lorentz reps are incapable of supporting translations except trivially. 

The minimum upgrade has spin $(A,B)\oplus(C,D),$ the reducible direct sum of two irreducible reps.\cite{Lyubarskii,S1} And the spins must be 'linked': $C$ = $A \pm 1/2$ and $D$ = $B \pm 1/2.$ The two $\pm$ signs are not correlated; a given spin $(A,B)$ Lorentz rep can be upgraded to as many as four Poincar\'{e} reps with spin $(A,B)\oplus(C,D).$

For example, perhaps the simplest, certainly the one with the fewest components, is the Dirac 4-spinor rep of the Poincar\'{e} group that has a Lorentz subgroup with spin $(0,1/2)\oplus(1/2,0).$ Under the Lorentz group's rotations and boosts, two of the four components transform with spin $(0,1/2)$ and the other two components transform with spin $(1/2,0).$ The translation subgroup involves all four components.

For matrix reps there are either Type I translation subgroups with spin $(C,D)$ terms added to $(A,B)$-spin quantities while leaving $(C,D)$ quantities invariant or, visa versa, Type II with $(C,D)$ quantities added to $(A,B)$ quantities. Like coordinate translations, both types are inhomogeneous: the added quantities do not depend on what they are being added to.

A ``choose or loose'' situation presents itself. There is apparently just one translation subgroup allowed by our spacetime experiences. By definition, the Poincar\'{e} group has just one translation subgroup.%Correspondingly, a Poincar\'{e} representation is defined to have just one translation subgroup.  For matrix reps there are either Type I translation subgroups: spin $(C,D)$ terms are added to $(A,B)$-spin quantities while leaving $(C,D)$ quantities invariant or Type II: $(C,D)$ quantities change by adding $(A,B)$ quantities. For those matrix Lorentz reps that can represent translations, the symmetry between $(A,B)$ and $(C,D)$ is necessarily broken when the Poincar\'{e} group is represented. 

While it is important to connect the theory with experimental results, those experiences rely on the behavior of particles that are well-described by quantum mechanics. Since quantum mechanics sometimes defies notions based on everyday experiences and yet must be consistent with them, it may be that keeping both translation subgroups can be used to describe behavior beyond, yet including, experiences with moving lab equipment and duplicating experimental results. 

So let's preserve the symmetry between the Lorentz spin reps $(A,B)$ and $(C,D)$ in this article just because symmetry is sometimes powerful and keep both Type I and Type II as dual-translations. The structure has a matrix rep of the Lorentz group as the foundation upon which ``dual-translations'' are constructed.

Closure is a problem. A group of transformations must include the results of successive transformations.  We show that in no circumstance, except trivially, does applying, say, a Type I translation followed by a Type II translation yield a transformation that is a combination of rotations, boosts, and the dual translations. New transformations result. 

Imagine what could happen if these new transformations combined with original transformations to produce even more new transformations, the process spiraling out of control.  In this article, excessive expansion of the Lorentz group by adding new transformations to maintain closure is deemed unsatisfactory. We seek spin combinations $A,B,C,D$ that produce satisfactory expansions of the Lorentz group.

We find that only the Dirac 4-spinor formalism, the spin $(1/2,0)\oplus(0,1/2)$ Lorentz rep with both dual-translations is closed by including just one new transformation. This new transformation is well-behaved. It mixes well with the others so that the resulting group of rotations, boosts, dual translations and the one new transformation is closed. The group represented by the Dirac 4-spinor Lorentz group plus the dual-translations plus the one new transformation forms a famous group, the conformal group.

If there was another spin combination $(A,B)\oplus(C,D)$  that satisfied the Lie algebra of the conformal group it would have been uncovered by the derivation. A corollary follows: the Dirac 4-spinor formalism  is the only spin $(A,B)\oplus(C,D)$ representation of the conformal group.

Since electromagnetism is conformally invariant\cite{Cunningham,Bateman} and the conformal group is the minimal of the semisimple groups that have the Poincar\'{e} group as a subgroup,\cite{Leznov} and for other reasons, the conformal group has been well-studied. Much of the literature involves manifolds and is only of peripheral interest here. For matrix reps, it is well-known that the Dirac 4-spinor formalism can represent the conformal group.\cite{Dirac,Liu}  

The Dirac 4-spinor formalism is not the only way to get a matrix representation of the conformal algebra. The $15 \times 15$ matrices of the adjoint representation form another matrix rep of the conformal algebra, but the adjoint rep contains several irreducible Lorentz spin reps, and is not one of the reps, spin  $(A,B)\oplus(C,D),$ considered in this article.

The rest of the article is divided into four sections plus an Appendix. The Appendix displays some well-known formulas for the generators of the Lorentz group and formulas for ``vector matrices'' needed for dual-translations.  Section 2 recalls aspects of the Lorentz group and clears the way for translations. The discussion centers on the Lie algebra of the group. The calculations in Sec. 3 complete the Lie algebra of the Lorentz group with dual-translations by providing expressions for the commutators of the dual-translation generators. 

The Lorentz rep plus dual-translations grows into a closed group by including new transformations as needed. The Lie algebra of the initial set of generators, Lorentz rep plus dual-translations, is shown in Sec. 3 to have several undesirable terms quadratic or higher-order in generators. Linearity is required of Lie algebras. 

In Sec. 4 it is shown how to mitigate or remove the undesirable terms by constraining the spins $(A,B)\oplus(C,D).$ Two of the four $A,B,C,D$ must vanish.  Thus only the Dirac formalism discussed in Sec. 5, i.e. the spin $(0,1/2)\oplus(1/2,0)$ Lorentz rep or its equivalent, allows a satisfactory outcome. It represents rotations, boosts and dual-translations in a group that is closed by the inclusion of just one additional generator that represents dilations.

\section{Rotations/Boosts; the Setup for Translations} \label{RotBooWith2Trans}

The Lorentz group of rotations/boosts in spacetime can be represented by finite-dimensional square matrices with complex-valued components. A matrix representation ${\bf{D}}_{ij}\left(\Lambda\right)$ of a Lorentz transformation $\Lambda$ is  generated by `angular momentum' matrices ${\bf{J}}^{\mu \nu},$ meaning  
\begin{equation} \label{expwJ}
{\bf{D}}_{ij}\left(\Lambda\right) = \left[\exp{(i\omega_{\rho \sigma}{\bf{J}}^{\rho \sigma}/2)}\right]_{ij} \; ,
\end{equation} %Weinberg QTF VolI p61 (2.4.3)
 where repeated indices are summed and the $\omega_{\rho \sigma}$ are parameters antisymmetric in $\rho \sigma$ that determine the Lorentz transformation $\Lambda.$ Indices $\mu,\nu,... \in$ $\{x,y,z,t\}$ = $\{1,2,3,4\}$ are for Minkowski coordinates with diagonal metric $\eta^{\mu\nu}$ = diag$(+1,+1,+1,-1).$ The notation for 3-space indices has $i,j,k,... \in$ $\{x,y,z\}$  = $\{1,2,3\}.$ 

One finds that the order of application of transformations is important; rotating about $x$ then about $y$ differs from rotating about $y$ first then $x$. For infinitesimal transformations, one arrives at the Lie algebra of the Lorentz group. The angular momentum matrices that generate spacetime rotations satisfy the well-known commutation relations
\begin{equation} \label{commJJ} i\left[{\bf{J}}^{\mu \nu },{\bf{J}}^{\rho \sigma }\right]  =  \eta^{\nu \rho }{\bf{J}}^{\mu \sigma }+\eta^{\mu \sigma }{\bf{J}}^{\nu \rho}-\eta^{\mu \rho }{\bf{J}}^{\nu \sigma}-\eta^{\nu \sigma}{\bf{J}}^{\mu \rho}\, ,
\end{equation} %20160415PoincareCommRules.nb
where $\left[{\bf{J}}^{\mu \nu },{\bf{J}}^{\rho \sigma }\right] := {\bf{J}}^{\mu \nu }\cdot {\bf{J}}^{\rho \sigma } - {\bf{J}}^{\rho \sigma } \cdot {\bf{J}}^{\mu \nu }$ and indices $\mu, \nu \in$ $ \{x,y,z,t \}$ = $ \{1,2,3,4 \}$ for  Minkowski coordinates in flat spacetime. The angular momentum matrices are antisymmetric in $\mu \nu,$  ${\bf{J}}^{\nu \mu}$ = $-{\bf{J}}^{\mu \nu},$ so just six can be independent. The study of angular momentum matrices has a long history and some well-known formulas for matrices satisfying (\ref{commJJ}) are displayed in the Appendix. 

Vectors transform under rotation/boosts in a way that preserves their scalar products. For a given matrix representation ${\bf{D}}_{ij}\left(\Lambda\right)$ of a rotation/boost $\Lambda,$ one can represent vectors as matrices ${\bf{V}}^{\mu}$ that transform both as a 2nd order tensor with the matrix transformation and as an ordinary vector with the transformation $\Lambda,$\cite{Tung} 
\begin{equation} \label{LambdaVector}
{\bf{D}}_{is}\left(\Lambda\right) {\bf{V}}^{\mu}_{s\bar{s}}{\bf{D}}_{\bar{s}j}^{-1}\left(\Lambda\right) = \left(\Lambda^{-1}\right)^{\mu}_{\sigma} {\bf{V}}^{\sigma}_{ij}
 \, .
\end{equation} %Weinberg QTF Vol 1 p214
For a transformation near the identity, $\Lambda_{\mu\nu}$ = $\eta_{\mu\nu} + \omega_{\mu\nu}$ with  $\mid \omega_{\mu\nu}\mid \; \ll 1 ,$  only low powers of ${\bf{J}}^{\rho \sigma}$ contribute significantly. Keeping just the terms linear in ${\bf{J}}^{\rho \sigma}$ implies, by (\ref{expwJ}) and (\ref{LambdaVector}), the commutation relation between an angular momentum matrix and a vector matrix,
\begin{equation} \label{commJV}
i\left[{\bf{J}}^{\mu \nu},{\bf{V}}^{\rho }\right] = \eta ^{\nu \rho}{\bf{V}}^{\mu}-\eta^{\mu \rho }{\bf{V}}^{\nu } \, .
\end{equation} %Weinberg QTF Vol 1 p214
These equations can be solved for the matrices ${\bf{V}}^{\mu}$ given the matrices ${\bf{J}}^{\mu \nu}.$

Suppose the ${\bf{J}}^{\mu \nu}$ angular momentum matrices represent an irreducible Lorentz representation with spin $(A,B),$ with $2A$ and $2B$ non-negative integers. When one solves (\ref{commJV}) for ${\bf{V}}^{\mu}$, one finds only the trivial solution ${\bf{V}}^{\mu}$ = 0. Irreducible Lorentz reps don't give vector matrices.

To have non-trivial vector matrices ${\bf{V}}^{\mu},$ one must combine at least two irreducible reps, $(A,B)\oplus(C,D)$ with ``linked,'' spins, where `linked' means \cite{Lyubarskii,S1}
\begin{equation} \label{ABlinkedCD} C = A \pm 1/2 \quad ; \quad D = B \pm 1/2 \, .
\end{equation} %Ref S1
Solutions of (\ref{commJV}) for the vector matrices ${\bf{V}}^{\mu}$ in each of the four cases of $\pm$ sign choices are given in the Appendix.

For the rep detailed in the Appendix, or by applying a suitable similarity transformation to some other rep, the angular momentum matrices and vector matrices are conveniently organized into block-matrix form,
\begin{equation} \label{JKForm2}
{\bf{J}}^{\mu \nu}  =  \begin{pmatrix}{\bar{\bf{J}}}^{\mu \nu}_{11} & 0 \cr 0 & \bar{{\bf{J}}}^{\mu \nu}_{22} \end{pmatrix} \quad  ;  \quad {\bf{V}}^{\rho}  =  \begin{pmatrix}0 & \bar{{\bf{V}}}^{\rho}_{12} \cr \bar{{\bf{V}}}^{\rho}_{21} & 0\end{pmatrix}\, .
\end{equation} %Appendix
The 11-block of the ${\bf{J}}^{\mu \nu}$ matrices has the spin $(A,B)$ angular momentum matrices and the 22-block has spin $(C,D)$ matrices. The 12- and 21-blocks of the vector matrices connect the two irreducible Lorentz reps.

Since (\ref{commJV}) is homogeneous in ${\bf{V}}^{\mu},$ there is an arbitrary overall scale factor. One can show that each block $\bar{{\bf{V}}}^{\rho}_{12}$ and $\bar{{\bf{V}}}^{\rho}_{21}$ has an independent scale factor $k_{12}$ and $k_{21}.$ Other than these two arbitrary constants, the components of ${\bf{V}}^{\mu}$ are determined by (\ref{commJV}) given ${\bf{J}}^{\mu \nu}.$

The generators of the translation subgroups are called ``momenta.'' The components of momentum form a vector, so the momentum is necessarily a vector matrix satisfying the commutation relation (\ref{commJV}). The momenta are thus determined except for the two arbitrary scale factors $k_{12}$ and $k_{21}$ of the 12- and 21-blocks of ${\bf{V}}^{\mu}$ in (\ref{JKForm2}).

Since a sequence of translations may be carried out in any order without changing the result, a translation subgroup is abelian. This implies the commutators of momenta components must commute. One can show that the momentum matrix must be entirely zero except for one of the off-diagonal blocks $\bar{{\bf{V}}}^{\rho}_{12}$ or $\bar{{\bf{V}}}^{\rho}_{21}$ in (\ref{JKForm2}). The choices are
\begin{equation} \label{V12andV21}
{\bf{P}}^{\rho}_{12}  =  \begin{pmatrix}0 & \bar{{\bf{V}}}^{\rho}_{12} \cr 0 & 0\end{pmatrix} \quad  {\mathrm{or}}  \quad {\bf{P}}^{\sigma}_{21}  =  \begin{pmatrix}0 & 0 \cr \bar{{\bf{V}}}^{\sigma}_{21} & 0\end{pmatrix}\, ,
\end{equation} %Ref S1
where the dual-momenta matrices are denoted ${\bf{P}}^{\rho}_{12} $ and $ {\bf{P}}^{\sigma}_{21}.$
By the off-diagonal form of the momentum matrices, the dual-translations applied to a quantity $\psi$ with spin $(A,B)\oplus(C,D),$ yield 
\begin{multline} \label{Ppsi}
\left[\exp{\left(-ix_{\rho}{\bf{P}}^{\rho}_{12}\right)} \right]\psi = \left({\bf{1}}-ix_{\rho}{\bf{P}}^{\rho}_{12}\right)\psi  =  \begin{pmatrix} \bar{\bf{1}}_{11} & -ix_{\rho}\bar{{\bf{V}}}^{\rho}_{12} \cr 0 & \bar{\bf{1}}_{22}\end{pmatrix} \begin{pmatrix} \bar{\psi}_{1} \cr \bar{\psi}_{2} \end{pmatrix} = \begin{pmatrix} \bar{\psi}_{1} -ix_{\rho}\bar{{\bf{V}}}^{\rho}_{12} \bar{\psi}_{2}  \cr \bar{\psi}_{2} \end{pmatrix}   \\
\\ 
\left[\exp{\left(-ix_{\rho}{\bf{P}}^{\rho}_{21}\right)}\right] \psi= \left({\bf{1}}-ix_{\rho}{\bf{P}}^{\rho}_{21}\right)\psi  =  \begin{pmatrix} \bar{\bf{1}}_{11} & 0 \cr -ix_{\rho}\bar{{\bf{V}}}^{\rho}_{21} & \bar{\bf{1}}_{22}\end{pmatrix} \begin{pmatrix} \bar{\psi}_{1} \cr \bar{\psi}_{2} \end{pmatrix} = \begin{pmatrix} \bar{\psi}_{1}   \cr \bar{\psi}_{2} -ix_{\rho}\bar{{\bf{V}}}^{\rho}_{21} \bar{\psi}_{1}\end{pmatrix} \, .
\end{multline} %Check on the fly
Therefore, the momentum ${\bf{P}}^{\mu}_{12}$ generates a Type I translation by adding $(C,D)$ quantities to $(A,B)$ quantities and ${\bf{P}}^{\mu}_{21}$ generates a Type II translation by adding spin $(A,B)$ quantities to spin $(C,D)$ quantities. 

By (\ref{commJV}), (\ref{JKForm2})  and (\ref{V12andV21}), the momenta  ${\bf{P}}^{\rho}_{12}$ and ${\bf{P}}^{\sigma}_{21}$ are vector matrices with commutators
\begin{equation} \label{commJV12JV21}
i\left[{\bf{J}}^{\mu \nu},{\bf{P}}_{12}^{\rho }\right] = \eta ^{\nu \rho}{\bf{P}}_{12}^{\mu}-\eta^{\mu \rho }{\bf{P}}_{12}^{\nu } \quad ; \quad i\left[{\bf{J}}^{\mu \nu},{\bf{P}}_{21}^{\sigma }\right] = \eta ^{\nu \sigma}{\bf{P}}_{21}^{\mu}-\eta^{\mu \sigma }{\bf{P}}_{21}^{\nu } \, .
\end{equation} %20160415PoincareCommRules.nb
One sees by (\ref{V12andV21}) that the components of each momentum commute, 
\begin{equation} \label{commPP}
\left[{\bf{P}}^{\mu}_{12},{\bf{P}}^{\nu}_{12}\right]  =  0 \quad  {\mathrm{and}}  \quad \left[{\bf{P}}^{\mu}_{21},{\bf{P}}^{\nu}_{21}\right]  =  0\, ,
\end{equation} %Check on the fly
so both generate abelian subgroups.

\section{Dual-Translation Commutators} \label{Dual}

In this article, the Lorentz generators ${\bf{J}}^{\mu \nu}$ and both sets of dual-translation ${\bf{P}}^{\rho}_{12}$ and ${\bf{P}}^{\sigma}_{21}$ are kept as generators. So far we have commutation relations in (\ref{commJJ}) for $ i\left[{\bf{J}}^{\mu \nu },{\bf{J}}^{\rho \sigma }\right]$ and in (\ref{commJV12JV21}) for ${\bf{J}}^{\mu \nu}$ with ${\bf{P}}^{\rho}_{12}$ and ${\bf{P}}^{\sigma}_{21}.$
Finding the commutation relations, $[{\bf{P}}^{\rho}_{12},{\bf{P}}^{\sigma}_{21}],$ completes the set of commutation relations for the generators $\{{\bf{J}}^{\mu \nu},{\bf{P}}^{\rho}_{12},{\bf{P}}^{\sigma}_{21}\}.$

  Unlike the previous commutation relations, the expressions for $[{\bf{P}}^{\rho}_{12},{\bf{P}}^{\sigma}_{21}]$ depend on the spins $(A,B)\oplus(C,D),$ with $(C,D)$ linked to $(A,B).$ It is convenient to have some linkage-dependent functions. Define $\epsilon_{C},\epsilon_{D},\phi,\theta,r ,$
\begin{equation}  \label{eCeDfCfD}
C = A + \epsilon_{C} /2 \; ; \; D = B + \epsilon_{D} /2 \; ; \; \phi := (1 + \epsilon)/2 \; ; \; \theta := (1 - \epsilon)/2 \; ; \; r := \epsilon_{C}/\epsilon_{D} \, ,
\end{equation} %Definitions
with $\epsilon_{C},\epsilon_{D} $ = $\pm 1$ and $\phi,\theta \in$ $\{0,1\}.$ With $\pm 1,$ we have $\epsilon_{C}/\epsilon_{D}$ = $\epsilon_{D}/\epsilon_{C}$ = $\epsilon_{C}\epsilon_{D},$ so there is some flexibility in the way the expressions are written. 

\begin{table}[htb!]
\caption{Linkage functions $\epsilon,$ $\phi,$ $\theta,$ and $r$ for all four cases.}
\label{eftrTable}
\centering
\vspace{0.2cm}
\begin{tabular}{c|ccc|ccc|c}  
       Case  & $ \epsilon_{C} \,  $ & $ \phi_{C} \,
        $  & $ \theta_{C} \,  $ & $ \epsilon_{D} \,
        $ & $ \phi_{D} \,
        $  & $ \theta_{D} \,  $ & $ r \,
        $\cr
        1 & $ -1\, $ & $ 0\, $ & $ 1\, $ & $ -1\, $ & $ 0\, $ & $ 1\, $ & $ +1\, $\cr
    2  & $ -1\, $ & $ 0\, $ & $ 1\, $ & $ +1\, $ & $ 1\, $ & $ 0\, $ & $ -1\, $\cr
    3 & $ +1\, $ & $ 1\, $ & $ 0\, $ & $ -1\, $ & $ 0\, $ & $ 1\, $ & $ -1\, $\cr
    4  & $ +1\, $ & $ 1\, $ & $ 0\, $ & $ +1\, $ & $ 1\, $ & $ 0\, $ & $ +1\, $\cr
\end {tabular}
\end{table}  %20160520CollectFormulas4Paper.nb

Given the formulas for ${\bf{P}}^{\rho}_{12}$ and ${\bf{P}}^{\sigma}_{21}$ in the Appendix, it is straightforward to find the commutator $[{\bf{P}}^{\rho}_{12},{\bf{P}}^{\sigma}_{21}].$ The nonzero parts of both ${\bf{P}}^{\rho}_{12}$ and ${\bf{P}}^{\sigma}_{21}$ are off-diagonal in (\ref{V12andV21}), so $[{\bf{P}}^{\rho}_{12},{\bf{P}}^{\sigma}_{21}]$ is block-diagonal.  One can  express the commutator as a sum of terms involving the the unit matrix ${\bf{1}},$ the angular momentum matrices ${\bf{J}}^{\mu \nu}$, and a new diagonal matrix ${\bf{D}},$ defined below. For the most general spins $(A,B)\oplus(C,D)$ linked in (\ref{eCeDfCfD}), as they must be, the result is
\begin{multline} \label{commPP1TO4}
i\left[{\bf{P}}_{12}^{\mu },{\bf{P}}_{21}^{\nu }\right] =\frac{{k_{21}}{k_{12}}\;i}{2\sqrt{2A+\phi_{C}}\sqrt{2B+\phi_{D}}}[ \\-\left(1+\epsilon_{C} + \epsilon_{D} + 2\epsilon_{C}A+2\epsilon_{D}B\right)\eta^{\mu \nu}{\bf{1}}   + 2{\it{i}} \left[ r+2\left(A+B \right)\left(r-1\right)-2\left(A-rB\right)^{2}\right]\eta^{\mu \nu} {\bf{D}} + \\ 
+{\it{i}} {\bf{J}}^{\mu \nu}+2\left(1+\epsilon_{C} + \epsilon_{D} + 2\epsilon_{C}A+2\epsilon_{D}B\right) {\bf{J}}^{\mu \nu}\cdot {\bf{D}} - {\it{i}}\left(\epsilon_{C} - \epsilon_{D} + 2\epsilon_{C}A-2\epsilon_{D}B\right)\epsilon^{\mu\nu}_{\;\;ab} {\bf{J}}^{ab}\cdot {\bf{D}} +\\
+2{\it{i}}\eta_{ab}\left( {\bf{J}}^{\mu a}\cdot {\bf{J}}^{\nu b} + {\bf{J}}^{\nu a}\cdot {\bf{J}}^{\mu b} \right)\cdot {\bf{D}}] \;,
\end{multline} %20160415PoincareCommRules.nb
where $\epsilon_{\mu\nu\rho\sigma}$ is the four-index antisymmetric symbol with $\epsilon_{xyzt}$ = +1.   

The new matrix ${\bf{D}}$ differs from a multiple of the unit matrix only by the signs of the 11- and 22-blocks. The matrix is written as ``${\bf{D}}$'' to conform with convention, do not confuse it with the others, the spin $D,$ the transformation matrix ${\bf{D}}_{ij}\left(\Lambda\right),$ or the spin matrix ${\bf{D}}^{i}.$  One defines  
\begin{equation} \label{Dmatrix}
{\bf{D}} =  \frac{i}{2}\begin{pmatrix}{\bf{\bar{1}}}_{11} & 0 \cr 0 & -{\bf{\bar{1}}}_{22}\end{pmatrix} \; ,
\end{equation} %20160415PoincareCommRules.nb
with ${\bf{\bar{1}}}_{11}$ and ${\bf{\bar{1}}}_{22}$  the unit matrices for the 11- and 22-blocks, respectively. 

The matrix ${\bf{D}}$ commutes with the angular momenta generators ${\bf{J}}^{\mu \nu}$ and has the following commutators with momenta generators,
\begin{equation} \label{commDV}
i\left[{\bf{D}},{\bf{P}}_{12}^{\rho }\right] = -{\bf{P}}_{12}^{\rho} \quad , \quad i\left[{\bf{D}},{\bf{P}}_{21}^{\rho }\right] = +{\bf{P}}_{21}^{\rho} \; .
\end{equation} %20160415PoincareCommRules.nb
It follows that ${\bf{D}}$ would make an excellent generator. We wait until later to promote it to the status of a generator.

The commutation relations (\ref{commJJ}), (\ref{commJV12JV21}), (\ref{commPP}), (\ref{commPP1TO4}), (\ref{commDV}) form a complete set for the Lie algebra of the 15 generators  $\{{\bf{J}}^{\mu \nu},{\bf{P}}^{\rho}_{12},{\bf{P}}^{\rho}_{21},{\bf{D}}\},$ if we include ${\bf{D}}$ prematurely.  A Lie algebra has commutators among generators that are linear combinations of generators. That is achieved in the next section.

	\section{Closure} \label{spinHALF}

	The goal is to have a closed group built on a representation of the Lorentz group with dual-translations.  The Lie algebra must be closed and the commutators between generators must be expressed as linear combinations of the generators. The remaining obstacle is the non-linearity evident in (\ref{commPP1TO4}). In this section, spins $A,B,C,D$ are found that make the commutation relations (\ref{commPP1TO4}) linear in generators.

The presence of products of generators in the expressions (\ref{commPP1TO4}) for the commutators $i\left[{\bf{P}}_{12}^{\mu },{\bf{P}}_{21}^{\nu }\right]$ may require defining new generators to make a linear combination of generators. If we allow the process to get out of hand, with commutators of new generators forming new products and begetting more new generators, the Lie group of all matrices with the appropriate number of components looms as a possible outcome. 

The character of the group as Lorentz plus dual-translations would be muddied if the result was the Lie group of all matrices of a given dimension. Fortunately, the admission of new generators can be curtailed well before it gets to be a burden.

One of the aberrant quantities in the expression for $[{\bf{P}}_{12}^{\mu},{\bf{P}}_{21}^{\nu}]$ commutators in (\ref{commPP1TO4}) is the following, 
\begin{equation} \label{JJD}
 {\bf{JJD}} := \eta_{ab}\left( {\bf{J}}^{\mu a}\cdot {\bf{J}}^{\nu b} + {\bf{J}}^{\nu a}\cdot {\bf{J}}^{\mu b} \right)\cdot {\bf{D}}\; ,
\end{equation} %defined in 20160520CollectFormulas4Paper.nb
which is symmetric in $\mu \nu$ and has terms quadratic in generators ${\bf{J}}^{\mu\nu}$ times the would-be generator  ${\bf{D}}.$   

We can use the representation described in the Appendix to rewrite the space-space, space-time and time-time components of ${\bf{JJD}}$ as functions of the basic spin matrices ${\bf{A}}^{i},{\bf{B}}^{i},$ ${\bf{C}}^{i},{\bf{D}}^{i}.$  With $\mu\nu$ = $xx,$ expression ${\bf{JJD}}$ (\ref{JJD})  becomes 
\begin{equation} \label{JJD1}
 \left({\bf{JJD}}\right)_{11}^{xx} =  \left[2\left(A(A+1) +B(B+1)\right)\bar{\bf{1}}_{11}+4\left(-{\bf{A}}^{x}\cdot {\bf{B}}^{x}+{\bf{A}}^{y}\cdot {\bf{B}}^{y}+{\bf{A}}^{z}\cdot {\bf{B}}^{z}\right)\right]\cdot {\bf{\bar{D}}}_{11} 
\end{equation}
\[\left({\bf{JJD}}\right)_{22}^{xx}  = \left[2\left(C(C+1) +D(D+1)\right)\bar{\bf{1}}_{22}+4\left(-{\bf{C}}^{x}\cdot {\bf{D}}^{x}+{\bf{C}}^{y}\cdot {\bf{D}}^{y}+{\bf{C}}^{z}\cdot {\bf{D}}^{z}\right)\right]\cdot {\bf{\bar{D}}}_{22}   \, ,
\] %20160528a.nb
with the 12- and 21-blocks vanishing. The other $\mu$ = $\nu$ expressions $\left({\bf{JJD}}\right)^{\mu\mu}$ have the same form and the expressions $\left({\bf{JJD}}\right)^{\mu\nu}$ with different $\mu$ and $\nu$ have just ${\bf{A}}^{x}\cdot {\bf{B}}^{y}\cdot {\bf{D}}$-type terms. 

Rather than include new generators that are quadratic or higher order in the original generators, let's choose spins to make the ${\bf{A}}^{i}\cdot {\bf{B}}^{j}$ terms vanish, thereby making ${\bf{JJD}}$ proportional to ${\bf{D}}.$ In order to remove these terms, one can choose $A$ = 0 or $B$ = 0 for the 11-block and either $C$ = 0 or $D$ = 0 for the 22-block. There is no other way to remove all the ${\bf{A}}^{i}\cdot {\bf{B}}^{j}$ terms in the ${\bf{JJD}}$ expressions and make $\left({\bf{JJD}}\right)^{\mu\nu}$ proportional to ${\bf{D}}.$ Let us remember this result: $A$ = 0 or $B$ = 0 for the 11-block and either $C$ = 0 or $D$ = 0 for the 22-block, and move on to another term.

Another quantity that needs attention in (\ref{commPP1TO4}) is the quantity 
\begin{equation} \label{eJD}
(\epsilon {\bf{JD}})^{\mu \nu} := \epsilon^{\mu\nu}_{\;\;\rho \sigma} {\bf{J}}^{\rho \sigma}\cdot {\bf{D}} \; ,
\end{equation} 
which is antisymmetric in $\mu\nu.$ Since the only generator that is antisymmetric in $\mu\nu$ is the angular momentum ${\bf{J}}^{\mu \nu},$  we can avoid introducing a new generator by  making $(\epsilon {\bf{JD}})^{\mu \nu} $  proportional to ${\bf{J}}^{\mu \nu}.$  By applying the formula (\ref{AB2JK}) from the Appendix to $(\epsilon {\bf{JD}})^{\mu \nu}$  with $\mu\nu$ = $xy$ in (\ref{eJD}), one gets 
\[ \left(\epsilon {\bf{JD}}\right)^{xy}  = \bar{k} {\bf{J}}^{xy} \]
\begin{equation} \label{eJDtoJ}
  \begin{pmatrix}{\bf{A}}^{z}-{\bf{B}}^{z} & 0 \cr 0 & -{\bf{C}}^{z}+{\bf{D}}^{z}\end{pmatrix} = \bar{k}\begin{pmatrix}{\bf{A}}^{z}+{\bf{B}}^{z} & 0 \cr 0 & {\bf{C}}^{z}+{\bf{D}}^{z}\end{pmatrix} \quad (\mu=x,\nu=y) \, ,
\end{equation}
for some nonzero proportionality constant $\bar{k}.$ From their definitions in the Appendix, ${\bf{A}}^{z}$ and ${\bf{B}}^{z}$ are linearly independent matrices. Likewise, ${\bf{C}}^{z}$ and ${\bf{D}}^{z}$ are linearly independent matrices. Thus, with $\bar{k}$ = +1, we must have spins $B$ = $C$ = 0 and with $\bar{k}$ = $-1$ one gets spin $A$ = $D$ = 0.

Thus ${\bf{JJD}}$ and $\epsilon {\bf{JD}}$ in (\ref{JJD}) and (\ref{eJD}) reduce to expressions linear in ${\bf{D}}$ and ${\bf{J}}^{\mu\nu}$ when one of the spins $A$ or $B$ is zero and one of the spins $C$ or $D$ is zero. Suppose $A$ = 0, then  $C$ = 1/2 by linkage (\ref{eCeDfCfD}), so $D$ must be zero by (\ref{eJDtoJ}), and $D$ = 0 implies $B$ = 1/2 by linkage. Thus assuming $A$ = 0 yields the Case 3 spins $(A,B)\oplus(C,D)$ = $(0,1/2)\oplus(1/2,0).$ Alternatively, if $B$ = 0, the same logic produces the Case 2 spins $(A,B)\oplus(C,D)$ = $(1/2,0)\oplus(0,1/2).$ By Table 1, we have, for both spin sets,
\begin{multline} \label{eJDJJD}
\hspace{3.5cm} - {\it{i}}\left(\epsilon_{C} - \epsilon_{D} + 2\epsilon_{C}A-2\epsilon_{D}B\right)\epsilon^{\mu\nu}_{\;\;ab} {\bf{J}}^{ab}\cdot {\bf{D}} = 3i{\bf{J}}^{\mu\nu}
\\
\\
2{\it{i}}\eta_{ab}\left( {\bf{J}}^{\mu a}\cdot {\bf{J}}^{\nu b} + {\bf{J}}^{\nu a}\cdot {\bf{J}}^{\mu b} \right)\cdot {\bf{D}} = 3i\eta^{\mu\nu} {\bf{D}} \, , \hspace{3cm} 
\end{multline} % 20160616SCRAP.nb
since, in (\ref{JJD1}), one has $2\left(A(A+1) +B(B+1)\right)$ = $2\left(C(C+1) +D(D+1)\right)$ = $3/2.$

The coefficients of the remaining terms in (\ref{commPP1TO4}) have the same values for both spin sets. By Table 1, one finds
\begin{multline}  \label{coeffsCase2and3}
\hspace{4cm} 1+\epsilon_{C} + \epsilon_{D} + 2\epsilon_{C}A+2\epsilon_{D}B = 0
\\
2{\it{i}} \left[ r+2\left(A+B \right)\left(r-1\right)-2\left(A-rB\right)^{2}\right] = -7i \; . \hspace{3cm} 
\end{multline} %Notes p1 6/29/16
The first of these  makes the ${\bf{1}}$ and ${\bf{J}}^{\mu \nu}\cdot {\bf{D}}$ terms drop out.

Collecting the results (\ref{eJDJJD}) and (\ref{coeffsCase2and3}) in (\ref{commPP1TO4}) shows that the two Lorentz reps $(1/2,0)\oplus(0,1/2)$ and $(0,1/2)\oplus(1/2,0)$ give the same commutation relation for $\left[{\bf{P}}_{12}^{\mu },{\bf{P}}_{21}^{\nu }\right],$
\begin{equation} \label{commV12V21A}
i\left[{\bf{P}}_{12}^{\mu },{\bf{P}}_{21}^{\nu }\right] = 2k_{12}k_{21}\left(  \eta^{\mu \nu} {\bf{D}} - {\bf{J}}^{\mu \nu}\right)  \; ,
\end{equation}
where the constants $k_{12}$ and $k_{21}$ are arbitrary.

The structure constants in (\ref{commV12V21}) appear to have an arbitrary scale factor $k_{12}k_{21}.$  However, since linear combinations of generators are generators, the Lie algebra of the linear combinations is equivalent  to the original Lie algebra. For example, such manipulations can make real structure constants all integers. Here, replacing ${\bf{P}}_{12}^{\mu }$ and ${\bf{P}}_{21}^{\nu },$
\begin{equation} \label{newPP}
{\bf{P}}_{12}^{\mu } \rightarrow {\bf{P}}_{12}^{\mu }/k_{21}\quad ; \quad {\bf{P}}_{21}^{\nu } \rightarrow {\bf{P}}_{21}^{\nu }/k_{12} \; ,
\end{equation} 
gives an equivalent Lie algebra but with 
\begin{equation} \label{k12k21}
k_{12}k_{21} = 1  \; .
\end{equation}
The replacements obey the commutation relation
\begin{equation} \label{commV12V21}
i\left[{\bf{P}}_{12}^{\mu },{\bf{P}}_{21}^{\nu }\right] = 2  \eta^{\mu \nu} {\bf{D}} - 2{\bf{J}}^{\mu \nu} \; , 
\end{equation}
where all the structure constants are integers.

Since the sum $\oplus$ is commutative, there exists a similarity transformation taking matrices for spin $(A,B)\oplus(C,D)$ to matrices $(C,D)\oplus(A,B).$  Thus there is just one solution for ${\bf{JJD}}$ and $\epsilon {\bf{JD}}$ reducing to ${\bf{D}}$ and ${\bf{J}}^{\mu\nu}$ occurring in the two equivalent spin sets  $(1/2,0)\oplus(0,1/2)$ and $(0,1/2)\oplus(1/2,0).$

The commutator (\ref{commV12V21}) would be linear in generators if ${\bf{D}}$ were a generator. It is finally time to include ${\bf{D}}$ in the list of generators. Now all commutators between generators are linear in the generators and the Lie algebra is closed for Case 2 spins  $(A,B)\oplus(C,D)$ = $(1/2,0)\oplus(0,1/2)$ and its equivalent, Case 3 spins $(A,B)\oplus(C,D)$ = $(0,1/2)\oplus(1/2,0).$  These two equivalent spin sets allow the Lorentz group with dual-translation subgroups to have a closed Lie algebra.

Collecting the commutators from (\ref{commJJ}), (\ref{commJV12JV21}), (\ref{commDV}), and (\ref{commV12V21}), one has the commutation relations of the Lie algebra. The generators of the Lie group are the $ 6+4+4+1 $ = 15 matrices $\{{\bf{J}}^{\mu \nu},{\bf{P}}^{\rho}_{12},{\bf{P}}^{\rho}_{21},$ ${\bf{D}}\},$ counting 6 independent ${\bf{J}}^{\mu \nu},$ since ${\bf{J}}^{\mu \nu}$ = $-{\bf{J}}^{\nu \mu}.$  The commutation relations are
	\[i\left[{\bf{J}}^{\mu \nu },{\bf{J}}^{\rho \sigma }\right]= \eta^{\nu \rho }{\bf{J}}^{\mu \sigma }+\eta^{\mu \sigma }{\bf{J}}^{\nu \rho}-\eta^{\mu \rho }{\bf{J}}^{\nu \sigma}-\eta^{\nu \sigma}{\bf{J}}^{\mu \rho}\] %20160415PoincareCommRules.nb
	\[i\left[{\bf{J}}^{\mu \nu},{\bf{P}}_{12}^{\rho }\right] = \eta ^{\nu \rho}{\bf{P}}_{12}^{\mu}-\eta^{\mu \rho }{\bf{P}}_{12}^{\nu } \quad , \quad i\left[{\bf{J}}^{\mu \nu},{\bf{P}}_{21}^{\rho }\right] = \eta ^{\nu \rho}{\bf{P}}_{21}^{\mu}-\eta^{\mu \rho }{\bf{P}}_{21}^{\nu }\] %20160415PoincareCommRules.nb
	\[i\left[{\bf{D}},{\bf{P}}_{12}^{\rho }\right] = -{\bf{P}}_{12}^{\rho} \quad , \quad i\left[{\bf{D}},{\bf{P}}_{21}^{\rho }\right] = +{\bf{P}}_{21}^{\rho}\] %20160415PoincareCommRules.nb
	\begin{equation} \label{allCOMMs}
i\left[{\bf{P}}_{12}^{\mu },{\bf{P}}_{21}^{\nu }\right] = 2 \left(\eta^{\mu \nu} {\bf{D}} - {\bf{J}}^{\mu \nu}\right)   \; ,
\end{equation}
and with all other commutators between generators vanishing.	The Lie algebra is closed.

Comparing the commutation relations (\ref{allCOMMs}) with those of the conformal group \cite{conformal} shows that the 15 matrices $\{{\bf{J}}^{\mu \nu},{\bf{P}}^{\rho}_{12},{\bf{P}}^{\rho}_{21},{\bf{D}}\}$ satisfy the Lie algebra of the conformal group. 

One recognizes the spin sets $(1/2,0)\oplus(0,1/2)$ and $(0,1/2)\oplus(1/2,0)$ as equivalent representations of the well-known Dirac 4-spinor formalism. Adding new generators to the 15 we already have would give something other than the conformal group.  

It should be mentioned that the $15\times 15$ adjoint rep built from structure constants in (\ref{commV12V21}) has angular momentum matrices ${\bf{J}}^{\mu \nu}$ that can be shown to reduce to spins $(1,0)\oplus(0,1)\oplus(1/2,1/2)\oplus(1/2,1/2)\oplus(0,0).$ It would be an interesting exercise to catalog the matrix reps of the conformal group. Perhaps it has been done.

Only spins of the form $(A,B)\oplus(C,D)$ are considered in this article. Since the only spins that prevent the number of generators increasing beyond 15 are those of the Dirac 4-spinor formalism, it follows that only the Dirac 4-spinor formalism provides a finite dimensional matrix rep of the conformal group with spins restricted to the form $(A,B)\oplus(C,D)$.

	\section{Dirac 4-spinor Formalism} \label{4spinor}
	
	%Most eqns checked in 20160616SCRAP.nb

	In this section, the Dirac 4-spinor formalism as a representation of the conformal algebra is discussed. Since the two spin sets $(0,1/2)\oplus(1/2,0)$ and $(1/2,0)\oplus(0,1/2)$ are related by a similarity transformation, only one is needed. Choose $(0,1/2)\oplus(1/2,0).$
	
	The formulas in the Appendix give vector matrices ${\bf{V}}^{\mu}$ for the spin $(0,1/2)\oplus(1/2,0)$  4-spinor  rep. One finds 
\begin{multline} \label{Vmu}
\hspace{2.5cm} {\bf{V}}^{x} =  \begin{pmatrix} 0 &  +k \sigma^{x} \cr + \sigma^{x}/k & 0 \end{pmatrix} \; ; \; {\bf{V}}^{y} =  \begin{pmatrix} 0 &  -k \sigma^{y}\cr - \sigma^{y}/k & 0  \end{pmatrix}   \, 
\\
\\
 {\bf{V}}^{z} =  \begin{pmatrix} 0 &  -k \sigma^{z}\cr - \sigma^{z}/k & 0  \end{pmatrix} \; ; \; {\bf{V}}^{t} =  \begin{pmatrix} 0 &  +k \sigma^{t}\cr -\sigma^{t}/k & 0  \end{pmatrix} \hspace{2.5cm}\, ,
\end{multline}
where $k_{12} \rightarrow k$ and  $k_{21} \rightarrow 1/k$ in view of the requirement (\ref{k12k21}) that $k_{12} k_{21}$ = 1. Define
\begin{equation} \label{Pauli}
\sigma^{x} = \begin{pmatrix}0 & 1 \cr 1 & 0  \end{pmatrix} \; ; \; \sigma^{y} = \begin{pmatrix}0 & -i \cr i & 0  \end{pmatrix} \; ; \;\sigma^{z} = \begin{pmatrix}1 & 0 \cr 0 & -1  \end{pmatrix} \; ; \;\sigma^{t} = \begin{pmatrix}1 & 0 \cr 0 & 1  \end{pmatrix} \; 
\end{equation}
The purpose of the Appendix is to give matrices for general linked spins $(A,B)\oplus(C,D)$ and it was not set up to give any particular representation of 4-spinors. Not surprisingly, the matrices ${\bf{V}}^{\mu}$ do not come out in any easily-recognized form, but one can show that they are indeed Dirac gamma matrices. 

Direct calculation shows that 
\begin{equation} \label{VVplusVV}
{\bf{V}}^{\mu}\cdot{\bf{V}}^{\nu}+{\bf{V}}^{\nu}\cdot{\bf{V}}^{\mu} = 2\eta^{\mu\nu}  {\bf{1}} \; .
\end{equation}
Since Dirac $\gamma$ matrices can be characterized as $4\times4$ matrices that obey (\ref{VVplusVV}), we take
\begin{equation} \label{VisGamma}
{\bf{V}}^{\mu} = \gamma^{\mu}   \; .
\end{equation}
 Equation (\ref{VVplusVV}) has many consequences. For example, (\ref{VVplusVV})  implies that $\gamma^{x}\cdot\gamma^{x}$ = ${\bf{1}}$ and $\gamma^{x}\cdot\gamma^{y}$ = $-\gamma^{y}\cdot\gamma^{x}$ and such relationships limit the number of independent matrix products of gammas to 16.\cite{Messiah}

If one wishes, one can find a similarity transformation to take the gamma matrices ${\bf{V}}^{\mu}$ into any other set. For example the similarity transformation matrix $S,$
\begin{equation} \label{S}
S = \begin{pmatrix}  -i\sigma^{x}/k & 0 \cr 0 & \sigma^{x}\end{pmatrix} \; ,
\end{equation}
takes the gamma matrices $\gamma^{\mu}$ = ${\bf{V}}^{\mu}$ in (\ref{Vmu}) to a more conventional rep.\cite{W4spinors} One has
\begin{equation} \label{SimGammaSim}
\tilde{\gamma}^{\mu} = S\cdot \gamma^{\mu}\cdot S^{-1} =  \left\{\begin{pmatrix} 0 & -i \sigma^{i} \cr i\sigma^{i} & 0\end{pmatrix},\begin{pmatrix} 0 & -i \sigma^{t} \cr -i\sigma^{t} & 0\end{pmatrix} \right\}\; .
\end{equation}
We continue with the gamma matrices from the Appendix, the $\gamma^{\mu}$ = ${\bf{V}}^{\mu}$ in (\ref{Vmu}).

 Once the gammas are known, one can find the angular momentum matrices ${\bf{J}}^{\mu\nu}$ by
  \begin{equation} \label{JfromGamma}
{\bf{J}}^{\mu\nu} = -\frac{i}{4}\left(\gamma^{\mu}\gamma^{\nu}-\gamma^{\nu}\gamma^{\mu}\right)  \; .
\end{equation}
This gives the same result as applying the formulas in the Appendix.

Define $\gamma^{5}$ as usual,
\begin{equation} \label{g5}
\gamma^{5} = i\gamma^{t}\gamma^{x}\gamma^{y}\gamma^{z} = \begin{pmatrix}  -\sigma^{t} & 0 \cr 0 & \sigma^{t}\end{pmatrix} = 2i{\bf{D}}\; ,
\end{equation}
by (\ref{Dmatrix}). Thus, one sees that the equation
\begin{equation} \label{DfromGamma5}
{\bf{D}} = -\frac{i}{2}\, \gamma^{5} \; ,
\end{equation}
determines ${\bf{D}}.$ 

Then projection matrices $\left({\bf{1}}\pm\gamma^{5}\right)/2$ can be used,
 \begin{equation} \label{ProjectThePP}
{\bf{P}}^{\mu}_{12} = \frac{1}{2}\left({\bf{1}}-\gamma^{5}\right)\cdot \gamma^{\mu} \quad ; \quad {\bf{P}}^{\mu}_{21} = \frac{1}{2}\left({\bf{1}}+\gamma^{5} \right)\cdot \gamma^{\mu} \; ,
\end{equation}
 to bring out the momentum matrices.

Now suppose one chooses a different set of gamma matrices such as ${\gamma^{\mu}}^{\prime}$ = $S^{\prime}\cdot \gamma^{\mu}\cdot {S^{\prime}}^{-1},$ where $S^{\prime}$ is a similarity transformation matrix, i.e. a matrix with $\det{S^{\prime}}$ = 1. Then the generators ${{\bf{J}}^{\mu\nu}}^{\prime},$ ${{\bf{P}}^{\mu}_{12}}^{\prime},$ ${{\bf{P}}^{\mu}_{21}}^{\prime},$ and ${{\bf{D}}}^{\prime}$ can be determined from   (\ref{JfromGamma}), (\ref{DfromGamma5}) and (\ref{ProjectThePP})  with $\gamma^{\mu} \rightarrow$ $\tilde{\gamma}^{\mu}.$ Since the commutation relations are invariant under similarity transformations, these generators satisfy the Lie algebra of the conformal group (\ref{allCOMMs}).

\begin{appendix}

\section{Appendix} \label{AppA}

The Appendix describes the rep used in this article to derive formulas and check results. The formulas are written for spacetime described with Minkowski coordinates for a signature +2 diagonal metric $\eta^{\mu\nu}$ = diag$(+1,+1,+1,-1).$  The 4-vector notation has Greek letters $\mu,\nu,... \in$ $\{x,y,z,t\}$ = $\{1,2,3,4\}.$ The spatial components of vectors and tensors are denoted by Roman indices $i,j,k,... \in$ $\{x,y,z\}$  = $\{1,2,3\}.$   
%I prefer the signature -2 metric and time as index 0.

{\bf{Spin Matrices.}} The formulas for rotation/boost matrices are standard and widely available.\cite{W+,Tung2} To represent the rotation/boost group for spin $(A,B),$ one can define matrices $({\bf{A}}^{x},{\bf{A}}^{y},{\bf{A}}^{z})$ and $({\bf{B}}^{x},{\bf{B}}^{y},{\bf{B}}^{z})$ with components 
	\begin{equation} \label{MatricesSpinA}
{\bf{A}}^{\pm \, \bar{a}\bar{b}}_{ab} = {\bf{A}}^{x \, \bar{a}\bar{b}}_{ab} \pm {\bf{A}}^{y \, \bar{a}\bar{b}}_{ab} = \sqrt{(A\pm a)(A\mp a+1)}\; \delta^{\bar{a}}_{a\mp 1}\delta^{\bar{b}}_{b}  \quad ; \quad {\bf{A}}^{z \, \bar{a}\bar{b}}_{ab} = a \, \delta^{\bar{a}}_{a}\delta^{\bar{b}}_{b} \; ,
\end{equation}
	\begin{equation} \label{MatricesSpinB}
{\bf{B}}^{\pm \,\bar{a}\bar{b}}_{ab} = {\bf{B}}^{x \, \bar{a}\bar{b}}_{ab} \pm {\bf{B}}^{y \, \bar{a}\bar{b}}_{ab} = \sqrt{(B\pm b)(B\mp b+1)}\; \delta^{\bar{a}}_{a}\delta^{\bar{b}}_{b\mp 1}  \quad ; \quad {\bf{B}}^{z \, \bar{a}\bar{b}}_{ab} = b \, \delta^{\bar{a}}_{a}\delta^{\bar{b}}_{b} \; ,
\end{equation}  %Weinberg QTF VolI p230 (5.6.16,17) & notes p1 6/28/16.
where $2A$ and $2B$ are positive integers and the indices range from $-A$ to $+A$ for $a, \bar{a} \in$ $\{-A,-A+1,...,A-1,+A\},$ and from $-B$ to $+B$ for $b, \bar{b} \in$ $\{-B,-B+1,...,B-1,+B\},$ producing $N_{11}\times N_{11}$ square matrices with dimension $N_{11}$ = $\left(2A+1\right)\left(2B+1\right)$. Each set of matrices ${\bf{A}}^{k }$ and ${\bf{B}}^{k}$ with $k \in$ $\{1,2,3\}$ = $\{x,y,z\}$ together with ordinary matrix multiplication and addition makes a representation of the rotation group in 3-space.

Two pairs of numbers $ab$ and $\bar{a}\bar{b}$ label a component of ${\bf{A}}^{k \; \bar{a}\bar{b}}_{ab}$ or ${\bf{B}}^{k\; \bar{a}\bar{b}}_{ab},$ $k \in$ $\{1,2,3\}$ = $\{x,y,z\}.$  This 'double index' notation is convenient for calculations, but cumbersome when displaying the matrices. To replace the double index $ab$ with a single index $i$ one can use
	\begin{equation} \label{2TO1index}
i = \left(A+a\right)\left(2B+1\right) +B+b +1 \; .
\end{equation}
with inverse
	\begin{equation} \label{1TO2index}
a={\mathrm{Floor}}\left[\frac{i-1}{2B+1}\right]-A \quad ; \quad b = i-\left(A+a\right)\left(2B+1\right)-B-1\; .
\end{equation}  %20160415PoincareCommRules.nb
The single index $i$ runs from $i$ = $1$ for double indices $(a,b)$ = $(-A,-B)$ to $i$ = $(2A+1)(2B+1) $ for $(a,b)$ = $(+A,+B).$ 

One can get a reducible representation of the Lorentz group of rotations/boosts with spin $(A,B)\oplus(C,D).$ Combine the matrices in (\ref{MatricesSpinA}) and (\ref{MatricesSpinB}) with similar ones for spins $C$ and $D$ to form rotation matrices $({\bf{J}}^{x},{\bf{J}}^{y},{\bf{J}}^{z})$ and boost matrices $({\bf{K}}^{x},{\bf{K}}^{y},{\bf{K}}^{z}),$
	\begin{equation} \label{AB2JK}
	{\bf{J}}^{k} = \begin{pmatrix}{\bf{A}}^{k} + {\bf{B}}^{k}  & 0 \cr 0 & {\bf{C}}^{k} + {\bf{D}}^{k} \end{pmatrix} \quad ; \quad  {\bf{K}}^{k} = -i\begin{pmatrix}{\bf{A}}^{k} -{\bf{B}}^{k} & 0 \cr 0 & {\bf{C}}^{k} -{\bf{D}}^{k}\end{pmatrix} \; ,
\end{equation} %Weinberg QTF Vol I p230 (5.6.6,7) & notes p1 6/28/16
in block-matrix notation with indices suppressed. The 11-blocks have dimension $N_{11}$ = $\left(2A+1\right)\left(2B+1\right)$ and the 22-blocks have dimension $N_{22}$ = $\left(2C+1\right)\left(2D+1\right)$ and the ${\bf{J}}^{k}$ and ${\bf{K}}^{k}$ matrices have dimension $N,$
\begin{equation} \label{N11N22N}
N = N_{11} + N_{22} = \left(2A+1\right)\left(2B+1\right) + \left(2C+1\right)\left(2D+1\right) \, .
\end{equation}
Each matrix ${\bf{J}}^{k}$ and ${\bf{K}}^{k}$ generates a rotation or boost in a 2-dimensional plane in spacetime. Labeling the matrix with the coordinate indices of the plane affected, one has
\begin{equation} \label{Jmunu}
	{\bf{J}}^{ij} = \epsilon^{ijk}{\bf{J}}^{k} \quad ; \quad {\bf{J}}^{it} = -{\bf{J}}^{ti} = {\bf{K}}^{i}   \; ,
\end{equation} % 20160415PoincareCommRules.nb
where $\epsilon^{ijk}$ is antisymmetric in its indices and $\epsilon^{xyz}$ = 1.  
%{{mzero, mJz, -mJy, mKx}, {-mJz, mzero, mJx, mKy}, {mJy, -mJx, mzero,    mKz}, {-mKx, -mKy, -mKz, mzero}};

%\pagebreak
{\bf{Vector Matrices}} 

Once the rotation/boost matrices are chosen, the commutation relations (\ref{commJV}) required of a vector matrix determine the vector matrices in terms of two arbitrary scale constants $k_{12}$ and $k_{21}.$  The formulas depend on the linkage of spins $(C,D)$ to $(A,B),$ $C = A \pm 1/2; D = B \pm 1/2.$  Condensing the formulas of  Ref. \cite{S1}  with the linkage functions $\epsilon,\phi,\theta$ in the text in (\ref{eCeDfCfD}) and Table 1, one finds that (\ref{commJV}) and (\ref{MatricesSpinA})-(\ref{Jmunu}) imply the following formulas for the vector matrices,
\begin{equation} \label{Vxpmiy12}   ({\bf{V}}_{12}^{x} \pm i{\bf{V}}_{12}^{y})_{ab}^{cd} = \left(\pm 1\right)^{\frac{\epsilon_{C}+\epsilon_{D}}{2}}  k_{12}\frac{\sqrt{A \mp \epsilon_{C} a +\phi_{C}}\sqrt{B \mp \epsilon_{D} b +\phi_{D}}}{\left(\theta_{C}\sqrt{2A}+\phi_{C}\right)\left(\theta_{D}\sqrt{2B}+\phi_{D}\right)} \enskip \delta_{a \mp 1/2}^{c} \delta_{b \mp 1/2}^{d}  \end{equation} %({\bf{V}}^{x\pm iy}_{12})_{ab}^{cd} =

\begin{equation} \label{Vzpmt12}   ({\bf{V}}_{12}^{z} \pm {\bf{V}}_{12}^{t})_{ab}^{cd} =   \left(\pm 1\right)^{\frac{\epsilon_{C}-\epsilon_{D}}{2}}  \epsilon_{C} k_{12}\frac{\sqrt{A \pm \epsilon_{C} a +\phi_{C}}\sqrt{B \mp \epsilon_{D} b +\phi_{D}}}{\left(\theta_{C}\sqrt{2A}+\phi_{C}\right)\left(\theta_{D}\sqrt{2B}+\phi_{D}\right)} \enskip \delta_{a \pm 1/2}^{c} \delta_{b\mp 1/2}^{d}   \end{equation}  %({\bf{V}}^{z\pm t}_{12})_{ab}^{cd} =

\begin{equation} \label{Vxpmiy21}   ({\bf{V}}_{21}^{x} \pm i {\bf{V}}_{21}^{y})_{cd}^{ab} =  \left(\pm 1\right)^{\frac{\epsilon_{C}+\epsilon_{D}}{2}}  k_{21}\frac{\sqrt{A \pm \epsilon_{C} a +\phi_{C}}\sqrt{B \pm \epsilon_{D} b +\phi_{D}}}{\left(\phi_{C}\sqrt{2A+1}+\theta_{C}\right)\left(\phi_{D}\sqrt{2B+1}+\theta_{D}\right)} \enskip  \delta_{c\mp 1/2}^{a} \delta_{d \mp 1/2}^{b} , \end{equation}  %({\bf{V}}^{x\pm iy}_{21})_{cd}^{ab} = 

\begin{equation} \label{Vzpmt21}  ({\bf{V}}_{21}^{z} \pm {\bf{V}}_{21}^{t})_{cd}^{ab} = -\left(\pm 1\right)^{\frac{\epsilon_{C}-\epsilon_{D}}{2}}  \epsilon_{C} k_{21}\frac{\sqrt{A \mp \epsilon_{C} a +\phi_{C}}\sqrt{B \pm \epsilon_{D} b +\phi_{D}}}{\left(\phi_{C}\sqrt{2A+1}+\theta_{C}\right)\left(\phi_{D}\sqrt{2B+1}+\theta_{D}\right)} \enskip \delta_{c \pm 1/2}^{a} \delta_{d\mp 1/2}^{b} \, ,    \end{equation} %({\bf{V}}^{z\pm t}_{21})_{cd}^{ab} =
where, as in the text in (\ref{JKForm2}), $\bar{{\bf{V}}}_{12}^{\mu}$ is the 12-block of ${\bf{V}}^{\mu}$ and $\bar{{\bf{V}}}_{21}^{\mu}$ is the 21-block of ${\bf{V}}^{\mu}.$ Single indices $(i,j)$ can be found from the double indices $(ab,cd)$ by (\ref{2TO1index}).

With vector matrices, both 12- and 21-blocks can be nonzero. Momentum matrices, however, must commute, and it follows that the dual-momenta matrices ${\bf{P}}^{\mu}_{12}$ and ${\bf{P}}^{\mu}_{12}$ can have just one off-diagonal block with nonzero components,
\begin{equation} \label{P12P21} {\bf{P}}^{\mu}_{12} =  \begin{pmatrix} 0 & \bar{{\bf{V}}}_{12}^{\mu}\cr 0 & 0 \end{pmatrix} \quad ; \quad {\bf{P}}^{\mu}_{21} =  \begin{pmatrix} 0 & 0\cr \bar{{\bf{V}}}_{21}^{\mu} & 0 \end{pmatrix} \, .
\end{equation}
 Equivalently, ${\bf{P}}^{\mu}_{12}$ is ${\bf{V}}^{\mu}$ with $k_{21}$ = 0 and ${\bf{P}}^{\nu}_{21}$ is ${\bf{V}}^{\nu}$ with $k_{12}$ = 0.

\end{appendix}

%\pagebreak

\end{document}